\documentstyle[11pt]{article}
\input epsf

\newlength{\awidth}
\newlength{\aheight}
\newlength{\uswidth}
\newlength{\usheight}
\def\preprint#1{\gdef\@preprint{#1}}
\def\bce{\begin{center}}
\def\ece{\end{center}}
\def\be{\begin{equation}}
\def\ee{\end{equation}}
\def\bea{\begin{eqnarray}}
\def\eea{\end{eqnarray}}

\newcounter{fignr}

\begin{document}
\baselineskip=.285in

\catcode`\@=11
\def\maketitle{\par
 \begingroup
 \def\thefootnote{\fnsymbol{footnote}}
 \def\@makefnmark{\hbox
 to 0pt{$^{\@thefnmark}$\hss}}
 \if@twocolumn
 \twocolumn[\@maketitle]
 \else \newpage
 \global\@topnum\z@ \@maketitle \fi\thispagestyle{empty}\@thanks
 \endgroup
 \setcounter{footnote}{0}
 \let\maketitle\relax
 \let\@maketitle\relax
 \gdef\@thanks{}\gdef\@author{}\gdef\@title{}\let\thanks\relax}
\def\@maketitle{\newpage
 \null
 \hbox to\textwidth{\hfil\hbox{\begin{tabular}{r}\@preprint\end{tabular}}}
 \vskip 2em \begin{center}
 {\Large\bf \@title \par} \vskip 1.5em {\normalsize \lineskip .5em
\begin{tabular}[t]{c}\@author
 \end{tabular}\par}
 \end{center}
 \par
 \vskip 1.5em}
\def\preprint#1{\gdef\@preprint{#1}}
\def\abstract{\if@twocolumn
\section*{Abstract}
\else \normalsize
\begin{center}
{\large\bf Abstract\vspace{-.5em}\vspace{0pt}}
\end{center}
\quotation
\fi}
\def\endabstract{\if@twocolumn\else\endquotation\fi}
\catcode`\@=12

\preprint{}
\title{\Large\bf Neutrinos Must be Tachyons
\protect\\[1mm]\  }
\author{\normalsize Eue Jin Jeong\\[1mm]
{\normalsize\it Department of Physics, The University of Texas at Austin, 
Austin, TX 78712}}

\maketitle

\def\gatij{\gamma^{ij}}
\def\gabij{\gamma_{ij}}
\def\ophi{\phi^{a}}
\def\hphi{\overline{\phi}^{a}}
\begin{center}
{\large\bf Abstract}\\[3mm]
\end{center}
\indent\indent
\baselineskip=.285in
The negative mass squared problem of the recent neutrino experiments[1-6] 
prompts us to speculate that, after all, neutrinos may be tachyons.  
There are number of reasons to believe that this could be the case.  
Stationary neutrinos have not been detected.  There is no evidence of right handed 
neutrinos which are most likely to be observed if neutrinos can be stationary.  
They have the unusual property of the mass oscillation between flavors which has not 
been observed in the electron families.  While Standard Model predicts the mass 
of neutrinos to be zero, the observed spectrum of $T_{2}$ decay experiments hasn't 
conclusively proved that the mass of neutrino is exactly zero.  Based upon these 
observations and other related phenomena, we wish to argue that there are too many 
inconsistencies to fit neutrinos into the category of the ordinary inside light cone 
particles and that the simplest possible way to resolve the mystery of the neutrino is 
to change our point of view and determine that neutrinos are actually tachyons. 
 
\noindent
\vspace{1cm}

\noindent

\newpage
\baselineskip=15pt

\pagenumbering{arabic}
\thispagestyle{plain}
\setcounter{section}{1}
\indent\indent
	According to the so far confirmed physical data and the Standard Model, it is 
	generally believed that neutrinos are massless and they travel at the speed of 
	light.  Actually this scenario fits close to most of the observed experimental 
	results.  The problem is that neutrino is a fermion and no other existing known 
	fermions travel at the speed of light.  And no other fermions violate the parity 
	at the same time having the problem of mass oscillation.    
	In this note, instead of attempting to define what neutrinos must be like, we 
	wish to propose a scenario that neutrinos must be tachyons by showing how the 
	behavior of a tachyon is physically consistent with the properties of neutrinos 
	observed so far. In one of their pioneering works, 
	Alan Chodos, Avi I. Hauser and V. Alan Kostelecky [7] have suggested in 1985 that 
	at least one of the neutrinos may be tachyons within the field theoretical framework.
	  The present report may be considered the generalization of their work largely based 
	  on the collective analysis of the widely scattered physical data concerning neutrinos.    
	
	Tachyon is a particle derivable from the energy invariance equation of special 
	relativity exactly the same way as positron was predicted from it.  In this 
	solution, tachyons must travel always faster than the speed of light and have 
	the negative mass squared.  The total relativistic energy is still positive 
	and it obeys Lorentz invariance.  And any gauge theory based on Lorentz 
	invariance will have it embedded in the theory.  In that sense, it is not 
	an unphysical particle any more than an electron or a positron.  In other 
	words, there is absolutely no physical law that prohibits the existence of 
	tachyons.  If we try to detect tachyons by a deliberate experiment, we 
	would know it by measuring its negative mass squared value or its faster 
	than the speed of light travel.  Unlike gauge particles which have integer 
	spin, tachyons have spin $1/2$ and the conjugate anti-particle has the 
	opposite sign of the mass and a lepton number may be assigned.  The Dirac matrix,
	 quantum mechanically extended from the relativistic 
	energy relation, predicts about electron only that it has spin $1/2$ and that 
	it has the real mass with the possible existence of the conjugate anti-particle.
	  The attachment of electric charge and the magnetic dipole moment for an 
	  electron is the result of the additional U(1) gauge symmetry that charged 
	  particles generally possess.  Tachyon as a member of the fermion 
	  family has all the physical properties similar to that of electron 
	  except for the mass, charge and its speed of travel and other properties 
	  directly associated with them.  In a similar manner, by the introduction 
	  of the gauge symmetry SU(2), tachyons acquire the properties of neutrinos 
	  which are governed by the weak interaction.
	  
	  Since tachyons always 
	  travel faster than the speed of light, they never exist in the rest state.  
	  The closer its speed gets to the speed of light, the more energy it carries.  
	  The infinite speed tachyon will have zero energy and the one with the 
	  speed of light will carry infinite amount of energy exactly the same 
	  way as ordinary matter particles would.  The absolute rest mass of a 
	  tachyon is defined by the energy of the tachyon when its speed is 
	  $\sqrt{2}$ times the speed of light.  Therefore, usual large energy 
	  $(>MeV)$ electron neutrinos are expected to have the velocity slightly 
	  larger than the speed of light which may also explain the earlier 
	  detection of neutrinos from the supernova prior to its visual confirmation, 
	  within the reasonable assumption that both the weak and the electromagnetic 
	  interactions occurred almost simultaneously at the time of the explosion.  

	It is not an ordinary particle in any sense of our physical intuition.  
	However, it is as physical as any other particles we observe every day 
	in the sense that Lorentz invariance and the gauge symmetry do not prohibit 
	the existence of it.  
	The fundamental physical laws of our universe have their strong foot hold 
	on Lorentz and gauge invariance for various particle interactions.  In fact, our 
	elementary particle physics has evolved around these two invariances on 
	top of the quantum principle.   
	Since it is generally expected that an explosion would indicate a sudden 
	increase of temperature and pressure inside the stellar object thereby ejecting 
	radiation and debris at the same time,  it is difficult to conclude that 
	when 99 percent of the stellar energy is released in the form of neutrinos 
	during the collapsing phase, the star can still have its original 
	form intact and wait for one more day for an explosion only to emit photons.  
	If we assume that the emission of neutrinos and lights have occurred 
	almost simultaneously (within 1 hour period) at the time of the explosion, 
	the early arrival 
	of neutrinos from the supernova SN 1987A would prove unequivocally 
	that the observed neutrinos are tachyons with its absolute 
	rest mass 1.3 keV.  
	This is based on the report by M. Aglietta et al. [8] where the light has 
	traveled the distance of 50kps for 170,000 years and the first 
	consecutive bursts of 5 neutrinos have an average energy of about 
	6.7 MeV with an advanced arrival time of 1.11 days prior to the light 
	which has been confirmed by the subsequent visual detection of the 
	supernova SN 1987A.  This experimental data itself is in 
	drastic contradiction to the assumption that the neutrino is massless 
	and travels at the speed of light.  

	One of the puzzling discrepancies in this line of argument is in 
	the fact that the observed upper limit of the absolute mass squared 
	values of neutrinos from the recent experiments [1-6] are much smaller 
	than the one predicted from the supernova.  However, their results have 
	been estimated from the assumption that neutrinos are inside light 
	cone particles which may have contributed to the smaller mass squared 
	value than the actual one.  The continuous, down to zero energy 
	spectrum of the neutrino makes it almost impossible to distinguish it 
	from the behavior of a photon, which explains the general consensus 
	that neutrino is a particle with zero mass which travels at the 
	speed of light.  Also, the imaginary neutrino mass would allow 
	the electron energy spectrum from $T_{2}$ decay experiments to have 
	the long end tail with no specific structure suggesting the non 
	existence of the real rest mass.  

	Determining the dark matter content using the neutrino mass also 
	has to be viewed from a different perspective since the rest mass 
	can no longer be real.  In this case, the average energy density of 
	the neutrinos emitted at the time of the big bang must be used instead.  
	Because of its weak interaction with matter and faster than the speed of 
	light travel, neutrino would not lose its energy during the expansion 
	and cool down phase of the universe since the reduction of its speed 
	would mean the spontaneous increase of its energy which violates the 
	conservation of energy principle.  Therefore, assuming that neutrinos 
	are actually tachyons, it is possible to consider neutrinos as a 
	strong candidate for the dark matter in the universe since the total 
	mass-energy of neutrinos contributes directly to the matter content.  
	The faster than the speed of light travel also makes it unlikely 
	that neutrinos may be captured near massive gravitating stars to 
	be detected.  It is more likely that neutrinos will be uniformly 
	distributed throughout the universe regardless of the local 
	gravitational force as the energy of the neutrino gets close to 
	zero.  The concept of the imaginary mass of neutrinos also makes 
	it more likely that they can exist in the inter-transitional states 
	between different flavors due to the phase oscillation.  Furthermore, 
	there is no way of detecting both helicities of a particle traveling 
	faster than the speed of light, which is an another refreshing evidence 
	that neutrinos are tachyons.      

	The investigation so far indicates that tachyons do not violate 
	the observed experimental data concerning neutrinos.  Our traditional 
	belief that a particle's rest mass should be real and it must travel 
	equal to or less than the speed of light doesn't help to solve the 
	mystery of neutrino.  On the historical side of the problem, the 
	positron has already set an example that the reality of the rest 
	mass has nothing to do with the reality of the particle itself.  
	If neutrinos are tachyons as they seem to be, it widely opens a 
	completely different perspective of the universe.  The main puzzling 
	consequence of this result may be that some type of energy can in fact 
	be transferred faster than the speed of light as far as the particles 
	that carry the energy have the negative mass squared.  Of course, the 
	problem is that energy in general form has no way to be transferred 
	faster than the speed of light at will without invoking weak interactions 
	unless other control mechanisms for such interaction are created.  
	It is also conceivable that the solar neutrino deficiency problem 
	which is an yet another mystery associated with neutrinos may also 
	be due to the fundamental assumption that neutrinos are normal 
	inside light cone particles in the standard solar model in addition to 
	their flavor oscillation problem.  
	
	As pointed out by H. Van Dam, Y. J. NG and L. C. 
	Biedenharn, [9] one may need to extend the usual framework 
	of field theory to include the nonscalar tachyonic particles.  However, we do not 
	believe this is an insurmountable barrier compared to the fundamental conceptual 
	difficulties one faces by putting neutrinos into the category of 
	the normal light cone particles in our field theory.  As indicated by Alan Chodos, Avi I. 
	Hauser and V. Alan Kostelecky, [7] the need for such an extension cannot be used 
	to exclude apriori existence of tachyons, rather it suggests that 
	more theoretical work is required to determine physically acceptable 
	modification of the usual non-tachyonic quantum field theory. 
	
	Based on the above 
	discussions we conclude that neutrinos are tachyons which have all the major physical 
	properties required for the observed neutrinos. 
	Before we explore other gauge symmetries for neutrinos, it must be carefully 
	investigated to see if all the observed experimental facts on neutrinos fit into 
	the predicted behavior of tachyons. The ultimate test of this scenario may 
	depend on at least 
	one order of magnitude higher resolution and statistics in the future 
	$T_{2}$ decay experiments as well as on the reanalysis of the past 
	experimental data in light of the tachyonic particle's point of view.

\def\hebibliography#1{\begin{center}\subsection*{References}
\end{center}\list
  {[\arabic{enumi}]}{\settowidth\labelwidth{[#1]}
\leftmargin\labelwidth	  \advance\leftmargin\labelsep
    \usecounter{enumi}}
    \def\newblock{\hskip .11em plus .33em minus .07em}
    \sloppy\clubpenalty4000\widowpenalty4000
    \sfcode`\.=1000\relax}

\let\endhebibliography=\endlist

\begin{hebibliography}{100}

\bibitem{B1} R. G. H. Robertson et al., Phys. Rev. Lett.{\bf 67}, 957(1991) 
\bibitem{B2} H. Backe et al., Nucl. Phys. B(Proc. Suppl.) {\bf 31}, 46(1993)
\bibitem{B3} W. Stoeffl, Bull. Am. Phys. Soc. {\bf 37}, 925(1992)
\bibitem{B4} H. Kawakami et al., Phys. Lett. B {\bf 256}, 105(1991)
\bibitem{B5} E. Holzschuh et al., Phys. Lett. B {\bf 287}, 381(1992)
\bibitem{B6} J. F. Wilkerson, Nucl. Phys. B (Proc. Suppl.) {\bf 31}, 32 (1993)
\bibitem{B7} A. Chodos, A. I. Hauser and V. Alan Kostelecky, Phys. Lett {\bf 150 B}, 431 (1985)
\bibitem{B8} M. Aglietta et al, ``Supernova 1987A in the Magellanic Cloud", 
(Proc. of the fourth Mason Astrophys. Workshop), 119 (1988)
\bibitem{B9} H. Van Dam, Y.J. NG and L.C. Biedenharn, Phys. Lett {\bf 158 B}, 227 (1985)

\end{hebibliography}
\end{document}